\documentclass[runningheads]{llncs}
\usepackage[T1]{fontenc}
\usepackage{graphicx}

\usepackage{amsmath}
\RequirePackage{hyperref}

\usepackage{amsmath}
\DeclareMathOperator*{\argmax}{arg\,max}

\usepackage{pifont}

\usepackage{booktabs,tabularx,ragged2e}
\usepackage{multirow}
\usepackage{graphicx}

\usepackage{soul}
\usepackage{bbm}
\usepackage{algorithm}
\usepackage[noend]{algpseudocode}
\usepackage[shortlabels]{enumitem}
\usepackage{amssymb}

\newcommand{\comment}[1]{}
\comment{multiple lines}

\usepackage{setspace}

\usepackage{tikz}
\usetikzlibrary{shapes.geometric, arrows}
\usetikzlibrary{positioning}
\usetikzlibrary{arrows.meta}

\begin{document}
\title{Redundancy-aware unsupervised ranking based on game theory - application to gene enrichment analysis}
\titlerunning{Redundancy-aware unsupervised ranking based on game theory}
\author{Chiara Balestra  \and
	Carlo Maj \and
	Emmanuel Müller \and
	Andreas Mayr
}
\authorrunning{Balestra et al.}
\institute{Department of Computer Science, TU  Dortmund, Germany \and Department of Medical Biometry, Informatics and Epidemiology (IMBIE),
	University Hospital of Bonn, Germany \and Center for Human Genetics, University Hospital of Marburg, Germany}

\maketitle              % typeset the header of the contribution
\begin{abstract}
    Gene set collections are a common ground to study the enrichment of genes for specific phenotypic traits. Gene set enrichment analysis aims to identify genes that are over-represented in gene sets collections and might be associated with a specific phenotypic trait. However, as this involves a massive number of hypothesis testing, it is often questionable whether a pre-processing step to reduce gene sets collections' sizes is helpful. Moreover, the often highly overlapping gene sets and the consequent low interpretability of gene sets' collections demand for a reduction of the included gene sets. \par
    Inspired by this bioinformatics context, we propose a method to rank sets within a family of sets based on the distribution of the singletons and their size. We obtain sets' importance scores by computing Shapley values without incurring into the usual exponential number of evaluations of the value function. Moreover, we address the challenge of including a redundancy awareness in the rankings obtained where, in our case, sets are redundant if they show prominent intersections. \par
    We finally evaluate our approach for gene sets collections; the rankings obtained show low redundancy and high coverage of the genes. The unsupervised nature of the proposed ranking does not allow for an evident increase in the number of significant gene sets for specific phenotypic traits when reducing the size of the collections. However, we believe that the rankings proposed are of use in bioinformatics to increase interpretability of the gene sets collections and a step forward to include redundancy into Shapley values computations.

    \keywords{game theory  \and redundancy reduction \and unsupervised ranking  \and gene set analysis}

\end{abstract}

\section{Introduction}

\comment{

    %               about references
    
    %        N O T    I N    T H E    P A P E R 
    
    \begin{itemize}
        \item \textcolor{red}{\cite{belinky_pathcards_2015} they use Jaccard rate to compare the pathways, they want to minimize redundancy among pathways n different sources (so the goal is different); but, they prove that using jaccard rate is legitimate as it works as well as some more sophisticated method to compare pathways that they mentioned as well $\rightarrow$ this legitimate the use for jaccard rate as well}
        \item \textcolor{red}{\cite{van_iersel_presenting_2008} is only a visualization tool that allows for more interpretability}
        \item \textcolor{red}{\cite{doderer_pathway_2012} increase interpretability}
    \end{itemize} 
    
}

% _____________________________________________________________________________
%\begin{linenumbers}

When working with gene sets collections, one of the main challenges is the sheer size and the following low interpretability of the many gene sets belonging to the same collection. We refer to gene sets or \emph{pathways} as sets of genes deriving from a biological classification of genes with respect to chemical or biological functions; these are grouped in variably sized collections based on some prior biological~\cite{liberzon_molecular_2015} or chemical function. From this grouping derived collections that are naturally arbitrary in size and contain partly overlapping gene sets. The collections are often used for Gene Set Analysis GSA: A variety of methods that assess the enrichment of genes in different gene sets concerning a phenotype with the aim of identifying biological mechanisms potentially associated with it given a list of target genes found to be relevant for the analyzed trait. The size of the gene sets collections ranges among some hundreds to several thousand of gene sets; hence, the high number of performed hypothesis tests makes the correction for multiple testing being a major challenge~\cite{dudoit_multiple_2008,noble_how_2009}. In order to avoid heavy multiple testing corrections, a basic approach is to reduce the number of hypothesis tests to perform; hence, a potential solution is the incorporation of \emph{unsupervised} approaches to reduce the dimension of the collections before even considering a specific phenotype.

Cooperative game theory (CGT), introduced by Shapley~\cite{ref:shapley},  allows to fairly allocate resources among players. In the recent years, CGT found place in supervised feature selection where the scores obtained by Shapley values help discriminating among relevant and non-relevant features for the task at hand. Shapley values derive their success from the flexible and non-demanding definition of the \emph{value function} making them a tool easily applicable in various context, e.g., supervised feature selection, interpretable machine learning, allocation of resources among others~\cite{rozemberczki2022shapley,lundberg_unified_2017,cohen_feature_2007}. 

We introduce here a new method based on CGT to assign importance scores to sets within families of sets allowing for a reduction of the size of these families in an unsupervised manner. We use the Shapley values in order to rank the sets depending on how their elements are distributed throughout the whole family; we show that the Shapley values are positively correlated with the size of the sets and that they are not aware of intersections among them. We tackle both these problems adding a pruning criteria and get a new ranking of sets which shows no-correlation with the sets' sizes and low overlap among sets similarly ranked. To the best of our knowledge, this is the first time that an unsupervised redundancy-aware ranking of sets appears in the literature. In the gene sets collections' case study, we illustrate that our punished Shapley values affect the correlation among sizes of gene sets and their position in the rankings although not directly meant to solve this issue. The rankings show nice behaviour regarding the redundancy reduction and our results suggest that the switch to smaller gene sets collections does not affect the coverage of the genes. \par
Coming back to our initial problem setup, we use the gene sets collections as an application environment of the theoretical method proposed and focus on the obtained results regarding the reduction of the overlaps among pathways and the high coverage of the genes in the collections.

\section{Related work}

CGT led to applications in computer science in various contexts and applications; In particular, Shapley values have been extended to supervised and semi-supervised feature selection~\cite{cohen_feature_2007,pfannschmidt_evaluating_2016}, networks and security~\cite{van_campen_new_2017} and explainable machine learning~\cite{lundberg_unified_2017}. A recent paper by Rozemberczki et al.~\cite{rozemberczki2022shapley} summarizes the major applications of Shapley values in machine learning literature. Moreover, importance scores based on Shapley values have been adapted to study the interaction among genetic and phenotypic characteristics for gene sets prioritization analysis~\cite{lucchetti_shapley_2010,moretti_combining_2008}. Shapley values' exact computation on the other hand requires $2^N$ evaluations of a value function where $N$ is the number of players. The computational complexity of these scores makes their application unfeasible as soon as the number of players increases and, as a response, many approximation techniques of Shapley values appeared in the literature~\cite{van_campen_new_2017,castro_polynomial_2009}. The introduction of microarray games~\cite{moretti_class_2007} reduces the computational challenges of exact Shapley values' computation to polynomial time under the assumption that the game can be written using only binary relationships (i.e., \emph{anomalous} vs. \emph{normal}, \emph{in} vs. \emph{not in} among others). Sun et al.~\cite{sun_game_2020} made use of Shapley values to rank genes by their relevance, i.e., the individual genes' synergistic influence in a gene-to-gene interaction network.

Gene sets collections are families of pathways based on some prior biological knowledge~\cite{liberzon_molecular_2015}. These collections are by nature arbitrary in size and contain partly overlapping gene sets, consequently leading to high redundancy and low interpretability. As discussed in Stoney et al.~\cite{stoney_using_2018}, several methods to take the overlap among pathways into account as well as maximizing the coverage of the genes included in gene sets' collections have been proposed in the literature. The proposed methods include visualization tools of redundancy among pathways, merging gene sets based on their similarity and integrating gene sets into a non-redundant single and unified pathway~\cite{belinky_pathcards_2015,van_iersel_presenting_2008,doderer_pathway_2012}. In 2016, Frost et al.~\cite{frost_unsupervised_2016} developed two unsupervised rankings for gene sets based on hypothesis testing. However, these methods do not include any information on the interactions among pathways themselves. In contrast to other redundancy reduction methods proposed for the same scope, our aim is a ranking of the original pathways in gene sets collections using Shapley values, thus relying on theoretical properties of fair allocation of resources. 

One of the main application of gene sets collections is enrichment analyses, e.g., via Fisher test or GSEA algorithm~\cite{subramanian_gene_2005,mathur_gene_2018}, that tests for the potential over/under-representation of the analyzed genes in specifically biologically annotated gene sets. In this context, the number of statistical hypothesis tests performed equals the number of pathways within the gene sets collection thus the correction for multiple testing becomes a major challenge~\cite{dudoit_multiple_2008,noble_how_2009}. Different kind of corrections for multiple testing are known to preserve the type-I error~\cite{hochberg_sharper_1988,holm_simple_1979} but often lead to a loss of power (type-II error)~\cite{nakagawa_farewell_2004}. Among them we recall the Bonferroni correction and the false discovery rate FDR~\cite{benjamini_controlling_1995,benjamini_control_2001}, where the latter is less conservative. Another simple approach to avoid loss of statistical power is to reduce the number of tests. Limiting the number of tested pathways within a gene sets collection w.r.t. to a specific phenotype could lead to a bias (post-selection inference~\cite{berk_valid_2013}) while the incorporation of \emph{unsupervised} approaches to reduce the dimension of the gene sets collection before considering a specific phenotype could be a solution. If the reduction of the number of tests that need to be performed is independent of the phenotype and preserves the maximum amount of information contained in the gene set collection, i.e., an high coverage of the genes, the typically inflated type-I error due to pre-screening is avoided.

\section{Ranking sets in families of sets}

We introduce some basic notions of cooperative game theory and definitions that we use within this work. In the Appendix, we illustrate some of the game theoretical concepts introduced through a toy example to help the reader to familiarize with them. 

% _____________________________________________________________________________
 
%               COOPERATIVE GAME THEORY
%               definitions from classical game theory
%               literature; why it has a broad application

% _____________________________________________________________________________

\subsection{Cooperative Game Theory}

Game Theory was first formalized by L. Shapley in 1951~\cite{ref:shapley}. A cooperative game is a pair $(\mathcal{N}, v)$ where $\mathcal{N}$ is a finite set of players and $v: 2^{\mathcal{N}} \rightarrow \mathbb{R}_+$ is the so-called \emph{value function}. $v$ maps each subset (usually referred as \emph{coalition}) of players $\mathcal{T}\subseteq\mathcal{F}$ to a real non-negative number $v(\mathcal{T})$. A fundamental assumption is that the empty coalition has null value $v(\emptyset) = 0$ while it is common to work with normalized games, i.e., $v(\mathcal{N}) = 1$.  

% _____________________________________________________________________________

Each player can participate in the creation of $2^{{N}-1}$ coalitions where $N = |\mathcal{N}|$. Shapley values have been introduced  to fairly compare the roles of players within a cooperative game. They represent one of the possibilities of fairly dividing the payoff of the grand coalition $\mathcal{N}$ and the amount of resources which is allocated to each of the players is fair concerning the contribution of the player in any possible coalition within the game; moreover, it has been proved that they represent the only possible allocation of resources satisfying numerous properties, i.e., the \emph{dummy}, \emph{symmetry}, \emph{null-player} and \emph{efficiency} properties~\cite{ref:shapley}. In particular, the efficiency property states that the sum of the Shapley values over the set of all players equals $v(\mathcal{N})$.

The Shapley value of a player $i$ is the average of its \emph{marginal contributions}  $v(\mathcal{T}\cup\{i\})-v(\mathcal{T})$ across all possible coalitions $\mathcal{T}\subseteq\mathcal{N}$, i.e.,
\begin{equation}\label{eq:sv}
    \phi_i(v)= \sum_{\mathcal{T}\subseteq \mathcal{N}\setminus\{i\}} \frac{(N-t-1)!\ t!} {(N-1)!}\cdot (v(\mathcal{T}\cup\{i\})-v(\mathcal{T}))
\end{equation}
%\linenumbers
where $N = |\mathcal{N}|$ and $t = |\mathcal{T}|$. 
% _____________________________________________________________________________ 

In Equation~\eqref{eq:sv}, the value function needs to be computed $2^N$ times. Due to the exponential complexity, computational problems raise when the number of players increases. However, one particular class of games, the \emph{Sum-Of-Unanimity Games} (SOUG)~\cite{ref:shapley}, admits a polynomial closed-form solution (see the Appendix for details) and microarray games are a special case of SOUG games. 

% _____________________________________________________________________________ 

%               SOUG GAMES

%               why do we need them? bc they are the base on which 
%               the computation of microarray games is constructed. 

%               in particular, we get the matrix B, then construct the
%               corresponding SOUG game v* and then we compute the SV

% _____________________________________________________________________________ 

% _____________________________________________________________________________ 

%       C O M P A R I S O N     W I T H     B O N A S S I

% \textcolor{red}{In Bonassi et al. 2007, they compute the importance of genes given the samples in which the genes are abnormal. So, basically we have gene $g$, all samples, for each sample check if it abnormal in the gene $g$ and put zeroes and ones accordingly. They get an importance of genes. Instead we do: we have a pathway $P$, all genes in the gene set, for each gene check if it is present in the pathway $P$ and put zeroes and ones accordingly. Thus we get an importance of pathways.}

% _____________________________________________________________________________ 

\subsection{Microarray games} 
Let us consider $\mathcal{F}= \{P_1, \ldots, P_N\}$ a family of sets. We denote with 
\begin{equation*}
    G= \{g\in P_i\mid i\in\{1,\ldots, N\}\} = \bigcup_{i\in\{1,\ldots, n\}}P_i
\end{equation*} 
the elements belonging to at least one set $P_i$ and $M= |G|$. Starting from $\mathcal{F}$ and $G$, we build a binary matrix $B\in \{0,1\}^{N\times M}$ where $B_{ij} = 1$  if $g_j \in P_i$ and $B_{ij} = 0$ otherwise. Transposing the definition given by Moretti et al.~\cite{moretti_class_2007}, for each element $g_j\in  G$, we look at the set of sets in which $g_j$ is present. Given the matrix $B$ and its column $B_j$, we define the \emph{support of $B_j$} $\text{sp}(B_j)$ as the set 
\begin{equation*}
    \text{sp}(B_j) = \{i\in\{1, \ldots, N\}\mid\ B_{ij}=1\}.
\end{equation*}
The \emph{microarray game} is then defined as the cooperative game $(\mathcal{N}, v^*)$ where 
\begin{equation}\label{eq:shapley_microarray}
    v^*(\mathcal{T}) = \frac{|\Theta(\mathcal{T})|}{|G|} =  \frac{|\{g_j\in G\mid \text{sp}(B_j)\subseteq \mathcal{T} \text{ and } \text{sp}(B_j)\neq\emptyset\}|}{|G|}.
\end{equation}

Following Sun et al.~\cite{sun_game_2020} approach, the value function can be easily expressed in terms of a linear combination of unanimity games where each column is interpreted as a unanimity game. Using this formulation of the value function, the computation of Shapley values is reduced to polynomial time.

% _____________________________________________________________________________ 

\subsection{Computation of Shapley values and game definition}\label{sec:svdefinition}
Shapley values are a common solution to assign fair scores to players within a cooperative game. On the other hand, they show an intrinsic problem: redundant players get similar scores thus implying they are ranked in close positions. In the Appendix, the \emph{glove game} example clearly illustrates the problem. We integrate into Shapley value a redudancy-awareness concept to rank players taking possible overlapping among them into account; In particular, the player ranked at the $(i+1)$-th place should be the least overlapping possible with the first $i$-ranked players $\{P_1,\ldots, P_i\}$. In order to achieve such a redundancy aware ranking, we introduce punishments for players which are highly correlated to the ones previously ranked. 

Each set $P_i$ contain a variable number $M_i$ of elements, i.e., $P_i = \{g_1, \ldots, g_{M_i}\}$; the sets in $\mathcal{F}$ are arbitrarily large and can overlap. We construct a microarray game based on the binary matrix $B \in \{0,1\}^{N\times M}$ where $N=|\mathcal{F}|$ and $M = |\cup_{i=1}^NP_i|$. Each row of $B$ represents a set $P_i$ and $B_{ij} = 1$ if $g_j\in P_i$ while $B_{ij}=0$ if $g_j\notin P_i$, i.e., each column $i$ represents the partial ordering relationship of the element $g_j$ belonging to the set $P_i$.

Given a set $P_i\in\mathcal{F}$, the Shapley value of $P_i$ is computed following these two steps as proposed in~\cite{sun_game_2020}:
\begin{enumerate}
    \item from the matrix $B$, we get $\mathcal{A} = \{\text{sp}(B_j)\}_{j\in M}\subseteq\mathcal{P}(\mathcal{F})$. Each set in $\mathcal{A}$ represents the support of the corresponding element $g_j\in G$.
    \item each Shapley value is computed through the formula:  
    \begin{equation}\label{eq:shapley_definition}
        \phi(P_i) = \frac{1}{M}\cdot\sum_{j = 1}^M\left(\mathbbm{1}(P_i \subseteq\, \text{sp}(B_j))\cdot\frac{1}{|\text{sp}(B_j)|}\right).
    \end{equation}
\end{enumerate}

We use the computed Shapley values to order the $P_i$s: The higher the Shapley value of a set $P_i$, the more important the set is in the microarray game defined. The importance scores are a measure of the amount of elements $g$ contained in $P_i$ which are often contained in other sets as well. Each $\phi(P_i)$ is a real number in $[0,1]$ and from Shapley values' properties we know that $\sum_{j=1}^N \phi(P_j)=1$. However, the ranking of sets given by the Shapley values alone is not aware of possible \emph{overlapping} among players. 

In the Appendix, we report a toy example to illustrate the computation of Shapley values in our setup.

% _____________________________________________________________________________ 

\subsection{Definition of goals: redundancy and coverage}\label{sec:jaccard_definition}
Different kind of redundancies among players can appear in a game depending on their structure; We aim for a redundancy aware ranking goal in family of sets. Two sets are \emph{redundant} if they contain a high amount of shared elements, i.e., if they show a large overlap. We use as redundancy measure the \emph{Jaccard index}~\cite{jaccard_etude_1901}, i.e., given $A$ and $B$ two sets, their Jaccard index is $j(A,B) = \frac{|A\cap B|}{|A \cup B|}$. The Jaccard index is a real number in $[0,1]$ where $j(A,B) = 0$ if and only if $A\cap B = \emptyset$ and $j(A,B) = 1$ if and only if $A = B$. 

Although the goal is redundancy reduction, we do not want to compromise with the \emph{coverage} of $G$. We hereby define various types of punishments and compare them w.r.t. coverage and redundancy.

\begin{enumerate}
    \item \textbf{Redundancy} -- We assign the \emph{Jaccard rate} or \emph{Jaccard score} $\text{jac}(\mathcal{S})$ as redundancy score to a family of sets $\mathcal{S}$, i.e.,
    \begin{equation}\label{eq:jaccard_rate}
        \text{jac}(\mathcal{S}) = \frac{1}{|\mathcal{S}|(|\mathcal{S}|-1)}\sum_{P_i,P_j \in \mathcal{S}, P_i \neq P} j(P_i, P_j).
    \end{equation} 
    The Jaccard score $\text{jac}(\mathcal{S})$ represents the average Jaccard index among pair of sets in $\mathcal{S}$; it is a non-negative real number in $[0,1]$ and $\text{jac}(\mathcal{S})=0$ if and only if all couples of sets in $\mathcal{S}$ do not overlap. %To minimize the redundancy in a subset of pathways $\mathcal{S}$, the sets should have intersections as small as possible.
    \item \textbf{Coverage} -- We defined the finite family of sets $\mathcal{F}$ and $G$ the set of elements in the sets. The family $\mathcal{F}$ represents a coverage of $G$ since each element $g\in G$ is contained at least in one set $P_i\in\mathcal{F}$. Given $\mathcal{S}\subseteq \mathcal{F}$, the \emph{coverage of $G$ given by $\mathcal{S}$} $c_G(\mathcal{S})$ is the percentage of elements $g\in G$ that are included at least in one set in $\mathcal{S}$, i.e., $c_{G}(\mathcal{S}) = {|\cup_{P_i\in\mathcal{S}}P_i|\cdot}\frac{100}{|G|}$. There is an obvious trade-off between the coverage given by $\mathcal{S}$ and its dimension. In the case study, we will investigate the success of our methods in preserving the coverage of the entire set while reducing the dimension of $\mathcal{F}$.
\end{enumerate} 
The ranking given by Shapley values shows a clear tendency to rank bigger sets first. Moreover, these bigger sets are more likely to overlap. When looking for rankings with low redundancy, we will hence also affect this association between rankings' positions and sizes of sets.

%%%%%%%%%%%%%%%%%%%%%%%%%%%%%%%%%%%%%%%%%%%%%%%%%%%%%%%%%%%%%%%%%%%%%%%%%%%%%%%%%%

%                   D I F F E R E N T         P U N I S H M E N T S

%%%%%%%%%%%%%%%%%%%%%%%%%%%%%%%%%%%%%%%%%%%%%%%%%%%%%%%%%%%%%%%%%%%%%%%%%%%%%%%%%%

\subsection{Different punishments and different rankings}\label{sec:rankings_different}

\begin{figure}[t]
    \centering
    \includegraphics[width=\textwidth]{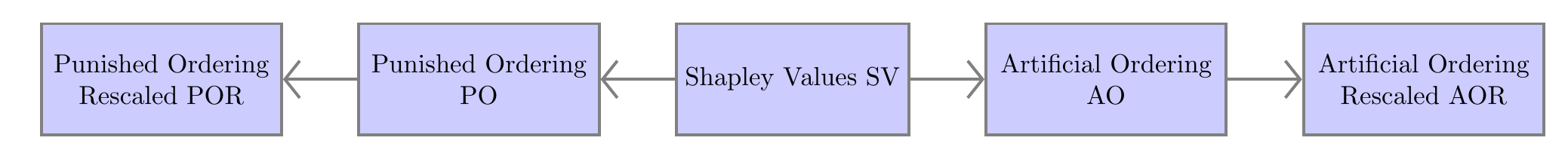}
    \vspace*{-5mm}
    \caption{\label{fig:genealogical_graph}Genealogical tree of the punishments' criteria.}
    %\vspace{-5mm}
\end{figure}

We introduce various punishment criteria. We provide a detailed comparison of the rankings obtained in Section~\ref{sec:case_study}: using the case study, we illustrate the properties of each of the punishments and how to select the most useful for the purpose at hand. An overview of the proposed methods highlights that
\begin{itemize}
    \item the proposed punishments are rather flexible and can be adapted to optimise specific properties, and
    \item there is not a perfect and unique choice fitting all applications.
\end{itemize}
The obtained rankings are constructed one on the top of the other as represented in Figure~\ref{fig:genealogical_graph}.

\begin{enumerate}
    
    % _____________________________________________________________________________ 
    
    %                                      S V
    
    \item \label{ord:SV}\textbf{Shapley values SV} -- we refer to the original ranking introduced by Equation~\eqref{eq:shapley_definition} as SV. This ranking favours larger sets and do not include any awareness of overlapping among sets. On the other hand, as the Shapley values tend to rank first larger sets, we expect an high coverage of $G$ when selecting even a small number of sets.
    
    % _____________________________________________________________________________ 
    
    %                                      P O
    
    \item \textbf{Punished Ordering PO} -- the Shapley values are used to obtain the first ranked set $\tilde{P}_1$, i.e., the one whose Shapley value is the highest. In the second step we re-compute all Shapley values for the not-yet ranked sets and we subtract from them  $j(\tilde{P}_1, P)$, i.e., the jaccard index among $\tilde{P}_1$ and the to-be-ranked set $P$. For each set $P$, the importance score $S_2(P)$ at the second step reads
    \begin{equation}
        S_2(P) =\phi_2(P)-j(\tilde{P}_1,P).
    \end{equation}
    The set $\arg\max_P S_2(P)$ obtains the second position and the process restarts. The punishment is increased in each step: in particular, after selecting the first $n$ sets, the score $S_{n+1}(P)$ obtained by the set $P$ (where $P$ has not been ranked yet at step $n+1$) is given by the following recursive formula
    \begin{equation*}
        \begin{array}{lcl}
            S_1(P) &= &\phi_1(P)\\

            S_{n+1}(P)&=& \phi_{n+1}(P) - \sum_{i=1}^n j(\arg\max_{\bar{P}} S_i(\bar{P}), P), \qquad \text{if } n\geq 0
        \end{array}
    \end{equation*}
    and, at the step $n+1$, the algorithm ranks the set $\tilde{P}_{n+1}=\arg\max_PS_{n+1}(P).$ 
     
    We underline that the Shapley values are re-computed after each iteration and  $\phi_n(\cdot)$ represents the Shapley value function at the $n$th iteration. The re-computation is necessary in order to satisfy the efficiency property, i.e., $\sum_{P_i} \phi_m(P_i) = 1$ for each $ m$ where the sum is computed over the sets not yet ranked. Moreover, the set $\mathcal{A}$ changes at each iteration implying a (possible) different ordering of the sets when the Shapley values are re-computed.
    
    % _____________________________________________________________________________ 
    
    %                                      P O R
    
    \item \textbf{Punished Ordering with Rescaling POR} -- POR ranking adds to PO a rescaling of the punishment: the term $ \sum_{i=1}^n j(\arg\max_{\bar{P}} S_i(\bar{P}), P)$ is re-scaled to the interval $[0, \max_P\{\phi_i(P)\}]$ at each iteration. 
    
    % _____________________________________________________________________________ 
    
    %                                      A O
    
    \item \textbf{Artificial Ordering AO} -- {we introduce an artificial set $AP$ updated at each iteration. The artificial set is initialized at step $1$ to $AP_1 = \argmax_{P}\phi_1(P)$.} At the $n$th iteration, $AP$ is updated the elements of the last ranked set
    \begin{equation*}
        AP_n = \cup_{i=0}^{n}\argmax_{P} S_i(P).
    \end{equation*}
     The importance score is defined as in PO, but punishing with a unique jaccard index with $AP_n$ instead of the single sets. In AO the punishment depends on the elements in $G$ that have been covered by the first $n$ ranked sets. The introduction of the artificial pathway $AP$ containing the elements that have been already covered avoids multiple punishments for the same overlapping elements.
    
    % _____________________________________________________________________________ 
    
    %                                    A O R
    
    \item \textbf{Artificial Ordering with Rescaling AOR} -- as in POR, the re-scaling is done for AO on the term $j(AP_n, P)$ to the interval $[0, \max_P\{\phi_i(P)\}]$.

\end{enumerate}
Note that each criterion order as first set the one with highest Shapley value. The orderings differ from each other from the second ranked set on.

The code for the proposed methods is publicly available on GitHub~\footnote{\url{https://github.com/chiarabales/geneset_SV}}. In the Appendix, we provide additional details on the construction of the punishments.
% The implementation code is available on \texttt{https://github.com/chiarabales/genesetSV}. 
\begin{algorithm}[!ht]\label{alg:PO}
	\caption{Punished Shapley Values}
	\begin{algorithmic}[1]
		\Procedure{}{$\mathcal{F}$, criteria, re-scaled = FALSE}
		\State take as input the family of sets $\mathcal{F}$
		\State $\tilde P_1 = \arg\max_{P}\{\phi_1(P)\}$
		\State $LS = \tilde P_1$
		\State $\mathcal{F} = \mathcal{F}\setminus \tilde P_1$
		\State $\text{initialize punish to } 0$
		%\While {$\mathcal{G} \neq \emptyset$}
		% ----------------------------------------------------
		\For {$i \in \texttt{range} (1, |\mathcal{F}|+1)$}
		\State $AP_i = \cup_{k < i}\tilde P_k$
		\For {$P\in \mathcal{F}$}
		
		\If {criteria $==$ PO}
		\State $\text{punish}(P) = \text{punish}(P) + j(LS, P)$
		\If {re-scaled == TRUE}
		\State  $\text{punish}(P) = \text{punish}(P)\cdot \frac{\max_k\{\phi_i(P_k)\}}{\max_k\{\text{punish}(P_k)\}}$
		\EndIf
		\State $S_i(P) = \phi_i(P) - \text{punish}(P)$

		% --------------------------------------------------------------------
        \ElsIf {criteria == APO}
		\State $\text{punish}(P) = j(AP_i, P)$
		\If {rescaled == TRUE}
		\State $\text{punish}(P) = \text{punish}(P)\cdot \frac{\max_k\{\phi_i(P_k)\}}{\max_k\{\text{punish}(P_k)\}}$
		\EndIf
		\State $S_i(P) = \phi_i(P) - \text{punish}(P)$
		\EndIf
	
	    \EndFor
	    \State $\tilde P_i = \arg\max_{P}\{S_i(P)\}$
	    \State $LS = \tilde P_i$
	    \State $\mathcal{F} = \mathcal{F}\setminus \tilde P_i$
	    \State $i++$
	    \EndFor
		\EndProcedure
		\item[]
		\Return{ordering $\tilde P$}
	\end{algorithmic}
\end{algorithm}
\setlength{\textfloatsep}{3pt}

%%%%%%%%%%%%%%%%%%%%%%%%%%%%%%%%%%%%%%%%%%%%%%%%%%%%%%%%%%%%%%%%%%%%%%%%%%%%%%%%%%

%                               C A S E    S T U D Y

%%%%%%%%%%%%%%%%%%%%%%%%%%%%%%%%%%%%%%%%%%%%%%%%%%%%%%%%%%%%%%%%%%%%%%%%%%%%%%%%%%

\section{A case study - gene sets and pathways ranking}\label{sec:case_study}

The implementation of our game-theoretic concept provides a promising new framework to reduce the dimension of gene sets collections while keeping a high coverage of genes. Gene sets collections $\mathcal{F}$ are sets of pathways and pathways $P_i$s are sets of genes $g_j$s. \par 
{We use these case study to illustrate that the proposed rankings} 
\begin{itemize}
    \item do not favour larger pathways 
    \item \emph{reduce the redundancy} among pathways
    \item maintain an \emph{high coverage} of the genes in the gene set collection.
\end{itemize}
Furthermore, we analyze the effects of the different punishments and descending pathways selections on the gene set enrichment analysis looking for the significance of pathways for different association traits. % Moreover, we investigate in the following sections how selecting a smaller amount of pathways influences the power of multiple statistical testing as the number of tests needed are lower than when considering the whole gene set. 

The experiments emphasize that the choice on which particular punishment to use is highly dependent on the goal and there is not a \textit{right} way of choosing. We summarize in Table~\ref{tab:summary} the properties of the different rankings. We present the results for four gene sets collections, i.e., the KEGG, CGN, CM and TFT LEGACY, and leave the details for the Appendix (Section~\ref{sec:data}) 

% _____________________________________________________________________________ 

% _____________________________________________________________________________ 

\subsection{Correlation with the pathways' dimensionality}\label{sec:bias_towards_dimensionality}

The Shapley value function assigns to a set $P\in\mathcal{F}$ a positive real number incorporating information on the distribution of its elements in the other sets of $\mathcal{F}$. This leads to a positive correlation with the sizes of pathways, i.e., sets with higher dimensions are more likely to get a higher Shapley value. In Table~\ref{fig:original_vs_bigsets}, Kendall's $\tau$ scores measure the ordinal association between the pathways' size and the position in the rankings. The table clearly displays that when ranking the pathways using SV, we tend to rank larger pathways first . Using AO and PO, this  effect is reversed in most of the gene sets collections; in particular, AO and PO rankings tend to first select small pathways while larger ones are ranked last in the orderings. In AOR and POR, there is no clear tendency of a correlation among the dimension of pathways and the position in the ranking. The rankings SV, PO and AO show similar behaviours across the different studied gene sets collections  while the re-scaled punishments show no clear tendency. \par
Our goal was to reduce the redundancy among pathways subsequently ranked. Hence, we indirectly affect the strength of the positive correlation among ranking position and their size as higher-dimensional pathways are more likely to show overlapping among them.

\begin{table}[!t]
	\small
    \centering
    \newcolumntype{C}{>{\centering\arraybackslash}X}
    \newcolumntype{R}{>1.5{\centering\arraybackslash}X}
    \begin{tabularx}{\textwidth}{@{}p{0.8cm}CCCCCCCC@{}}
    \toprule
    && {\textbf{SV}} & {\textbf{PO}} & {\textbf{POR}} & {\textbf{AO}} & {\textbf{AOR}}\\
    \midrule
    \textbf{KEGG} && 0.49 & -0.21 & 0.15 & 0.23 & 0.19 \\
    \textbf{CGN} && 0.43 & -0.51 & 0.019 & -0.702 & -0.041 \\
    \textbf{CM} && 0.759 & -0.531 & 0.019 & -0.702 & -0.041 \\
    \mbox{\textbf{TFT LEGACY} }&& 0.679 & -0.460 & 0.354 & -0.828 & 0.458 \\
    
    \bottomrule
    \end{tabularx}
    \caption{\label{fig:original_vs_bigsets} Kendall's $\tau$ coefficients measuring the correlation among the position in the ranking and the size of the gene sets.}
    \vspace{-2mm}
\end{table}
% _____________________________________________________________________________ 

\subsection{Redundancy elimination}\label{sec:redundancy_elimination}
\begin{figure}[t]
    \centering
    \includegraphics[width = \textwidth]{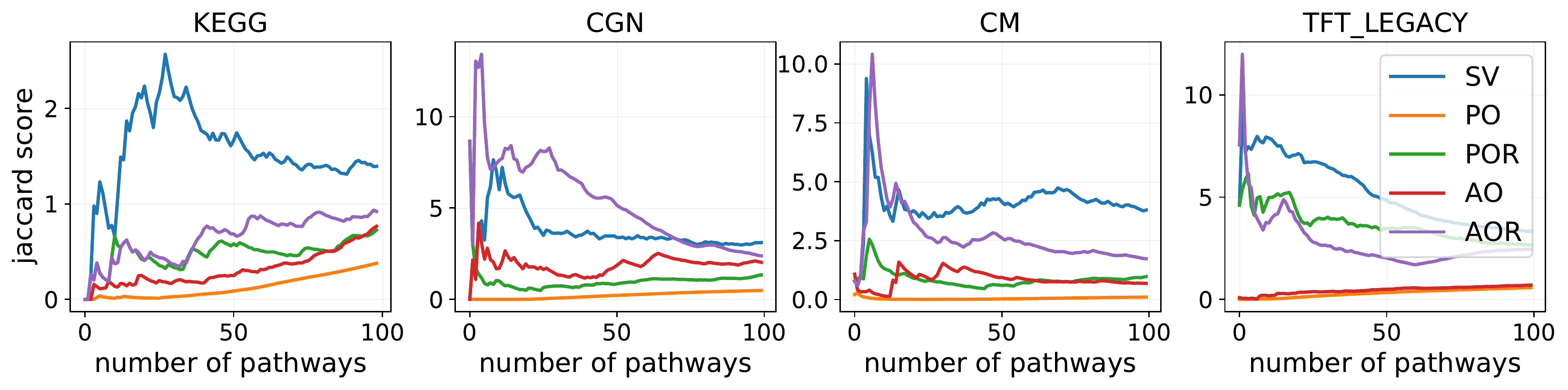}\\
    \vspace*{0.3cm}
    \small
    \centering
    \newcolumntype{C}{>{\centering\arraybackslash}X}
    \begin{tabularx}{\textwidth}{@{}p{0.8cm}CCCCCCCCCCCc@{}}
    \toprule
    & \multicolumn{3}{c}{\textbf{KEGG}} & \multicolumn{3}{c}{\textbf{CGN}} & \multicolumn{3}{c}{\textbf{CM}} & \multicolumn{3}{c}{\textbf{TFT LEGACY}}
    \\ \cmidrule(lr){2-4} \cmidrule(lr){5-7}\cmidrule(lr){8-10} \cmidrule(lr){11-13}
  	\% & 10 & 20 & 40 & 10 & 20 & 40 & 10 & 20 & 40 & 10 & 20 & 40 \\
    \midrule
    \textbf{SV} & {1.95} & {2.11} & {1.37} & 3.73 & {3.11} & {2.75} & {3.87} & {4.13} & {2.16} & {4.31} & {3.10} & {2.80} \\ 
    \textbf{PO} & \underline{0.02} & \underline{0.04} & \underline{0.21} & \underline{0.14} & \underline{0.42} & \underline{0.72} & \underline{0.02} & \underline{0.08} & \underline{0.21} & \underline{0.39} & \underline{0.65} & \underline{1.09}\\ 
    \textbf{POR} & {0.49} & 0.47 & 0.46 & {0.79} & {1.07} & 1.45 & {0.51} & 0.90 & 1.37 & 3.5 & 2.51 & 2.22\\ 
    \textbf{AO} & 0.15 & 0.19 & {0.38} & 1.21 & 1.97 & {1.36} & 1.19 & {0.80} & {0.63}  & {0.56} & {0.81} & {1.36}\\ 
    \textbf{AOR} & {0.49} & {0.45} & 0.77 & {5.82} & 2.97 & 2.20 & 2.56 & 2.00 & 1.25 & 1.70 & 2.39 & 2.43\\ 
    \bottomrule
    \end{tabularx}
    \vspace*{-0.2cm}
    \caption{\textbf{Redundancy reduction.} In the plots: average re-scaled Jaccard scores of sets of pathways ranked up to $j$-th position ($x$ axis). In each of the gene sets collection, we select up to $100$ pathways. In the table: re-scaled Jaccard rates of the first $10$, $20$ and $40\%$ of the gene sets; In each column, underlined text indicate the minimum Jaccard scores.}
    \label{fig:jaccard_rate}
    \vspace{2mm}
\end{figure}

The introduced punishments allow for rankings that minimize the overlaps subsequent ranked gene sets. We evaluate the redundancy reduction using the Jaccard score as defined in Equation~\eqref{eq:jaccard_rate}. To get comparable numbers throughout different gene sets collections, we re-scale the Jaccard scores to the \emph{maximum Jaccard index}, i.e., the maximum Jaccard index among any pairs of pathways within the gene set collection. In Figure~\ref{fig:jaccard_rate}, we plot the re-scaled Jaccard score as a function of the number of pathways. We first compute the rankings, then we select the respective first $n$ ranked gene sets and compute the Jaccard rate of the obtained set. The lower the Jaccard rate, the better the ranking performs in selecting non-overlapping pathways. In the Table~\ref{fig:jaccard_rate}, we compare the re-scaled Jaccard rates for some reference values. The rankings achieving the lowest Jaccard rates are the PO and the AO in (almost) all gene sets collections; PO and AO use strong punishments such that highly overlapping pathways with the first are ranked far from each other. Moreover, we note that the SV ordering is performing the worst in all but one case. 

% _____________________________________________________________________________ 

\subsection{Coverage of gene sets}\label{sec:coverage_of_genesets}

\begin{figure}[t]
    \centering
    \includegraphics[clip,width=\textwidth]{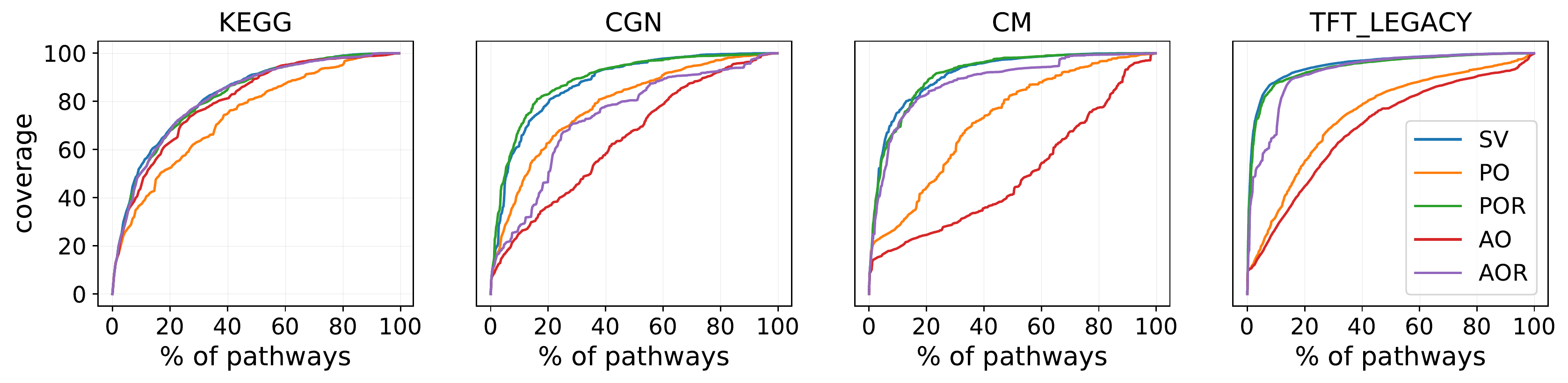}\\
    \vspace*{0.3cm}

    \scriptsize
    \small
    \centering
    \newcolumntype{C}{>{\centering\arraybackslash}X}
    \begin{tabularx}{\textwidth}{@{}p{0.8cm}CCCCCCCCCCCc@{}}
    \toprule
    & \multicolumn{3}{c}{\textbf{KEGG}} & \multicolumn{3}{c}{\textbf{CGN}} & \multicolumn{3}{c}{\textbf{CM}} & \multicolumn{3}{c}{\textbf{TFT LEGACY}}
    \\ \cmidrule(lr){2-4} \cmidrule(lr){5-7}\cmidrule(lr){8-10} \cmidrule(lr){11-13}
  	\% & 10 & 20 & 40 & 10 & 20 & 40 & 10 & 20 & 40 & 10 & 20 & 40 \\
    \midrule
    \textbf{SV} & \underline{52.7} & \underline{68.4} & \underline{86.2} & {61.3} & {79.3} & {93.4} & \underline{75.4} & {85.6} & {95.9} & \underline{88.2} & \underline{93.3} & \underline{97.0}\\ 
    \textbf{PO} & {37.0} & {52.4} & {74.7} & 42.6 & 62.7 & 81.6 & 28.0 & 44.2 & 73.2 & 32.3 & 55.1 & 78.6\\ 
    \textbf{POR} & {50.0} & {67.7} & 84.6 & \underline{68.0} & \underline{82.9} & \underline{93.5} & {68.9} & \underline{87.9} & \underline{96.2} & {87.5} & {91.8} & {96.1}\\ 
    \textbf{AO} & 43.7 & 63.0 & 81.2 & {25.0} & {36.3} & {57.7} & {19.1} & {24.5} & {35.6} & {27.4} & {44.7} & {70.7} \\ 
    \textbf{AOR} & {50.0} & 67.6 & {85.9} & 28.0 & 46.4 & 77.9 & {68.9} & 82.7 & 91.7 & 66.1 & 91.0 & {96.6} \\
    \bottomrule
    \end{tabularx}
    \vspace*{-2mm}
    \caption{\label{fig:cumulative}\textbf{Cumulative coverage of gene sets.} In the plots: the coverage of the genes as a function of the number of pathways included using the different rankings. In the Table: coverage of the genes when selecting respectively the $10$, $20$ and $40\%$ of the pathways. In each column, underlined text highlight the highest coverage.}
    \vspace{2mm}
\end{figure}

We investigate the ability of our methods to cover the genes using a limited amount of pathways which are ranked first. In Figure~\ref{fig:cumulative}, we plot the coverage of the genes {in percentage} when only considering the first $n$ ranked pathways. The SV ranking get a generally high coverage of genes in the gene sets collections. We note that the orderings SV, POR and AOR are clearly outperforming the rankings given by PO and AO. 
%Some recent work~\cite{belinky_pathcards_2015,van_iersel_presenting_2008,doderer_pathway_2012} underlined the necessity of covering the gene set using a minimum amount of pathways possible; some of the proposed methods modify the pathways in order to achieve high coverage and low redundancy among them. In contrast, we aimed to rank the pathways using importance scores that include a redundancy notion; we illustrate here how the introduced punishments behave concerning the coverage of the gene sets. 

Moreover, we compare the different rankings using some reference levels: In particular, the Table~\ref{fig:cumulative} gives insights on the proportions of the genes that can be covered using only a limited percentage of pathways. The high coverage achieved by SV is due to the correlation between the size of the pathways and their positions in the ranking; however, not the same can be argued about POR and AOR as we demonstrated that the correlation with the gene sets' sizes is reduced. Maximizing Shapley values while minimizing redundancy achieves outstanding performances in both cases. The lower performances of PO and AO are due to te fact that the punishments are generally harsh for overlapping gene sets; Hence, they select first small pathways which were ranked in the lowest positions by Shapley values alone. 

On the other hand, we observe that the rankings do not outperform the original SV ordering in covering the entire gene set; the advantages of the punished orderings are clear when considering that the performances of the newly proposed rankings are close to the original SV ranking while retaining a much smaller amount of redundancy and not preferring large pathways.

% _____________________________________________________________________________ 

\subsection{Statistical power}\label{sec:statisticalpower}
\begin{figure}[t]
    \includegraphics[width = \textwidth]{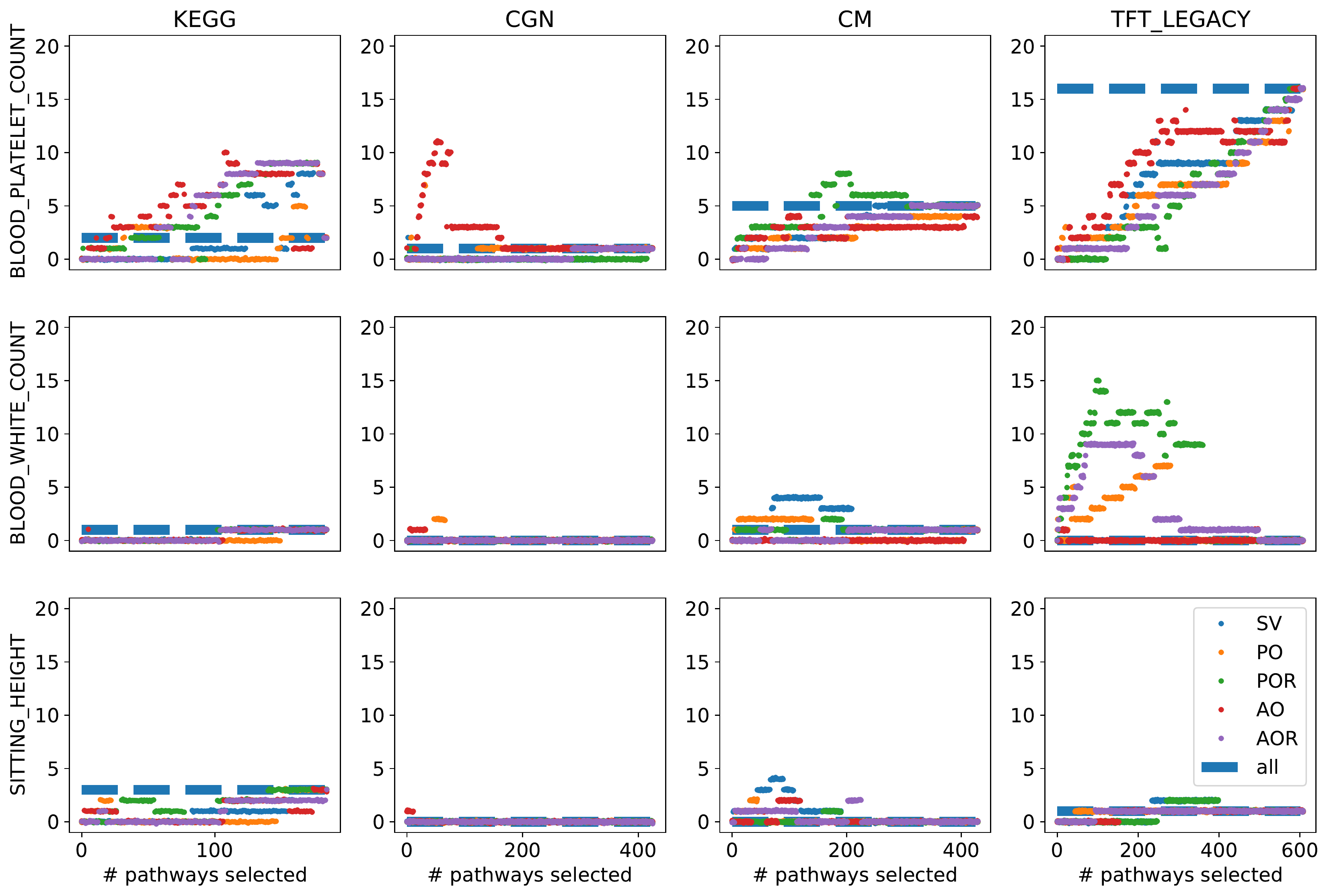}
    \vspace*{-5mm}
    \caption{\textbf{Increase in statistical power.} We show results on three selected association traits for the four gene sets. In each plot, the $x$-axis represents the number of pathways included in the multiple statistical testing while the $y$-axis represents the number of statistical significant pathways found. The different colours represent the different proposed orderings as in the legend. We used FDR correction for multiple testing and $\alpha = 0.05$.}
    \label{fig:statistical_power_increase}
    %\vspace{5mm}
\end{figure}
Finally, we evaluate how the proposed rankings behave with respect to gene set analysis. We use the rankings to obtain smaller gene sets collections and then test for associations with phenotypic traits. We show analysis when using the Fisher's exact test~\cite{fisher_logic_1935,agresti_introduction_2018}. We select the first $n$ ranked gene sets in each of the different rankings and correct the p-values using Bonferroni or FDR correction~\cite{benjamini_controlling_1995,dudoit_multiple_2008}. %We plot the number of significant pathways found after the correction for all the proposed rankings. 

In Figure~\ref{fig:statistical_power_increase} we illustrate the number of statistical significant gene sets founds for some association traits, i.e., \emph{blood platelet count}, \emph{blood white count} and \emph{sitting height}, w.r.t. the different gene sets collections; The plots refer to the FDR correction for multiple testing. The number of significant pathways in each of the gene sets collection are represented in each plot as a blue dashed line. We observe that the number of significant pathways found when limiting the number of tested pathways using the introduced rankings highly depends on the gene sets collection and the particular trait. In some settings using only a limited number of pathways may lead to a higher number of pathways reaching significance, but there are also settings when the number of significant pathways is reduced. This happens when significantly associated pathways are not ranked among the first, which obviously can take place when applying our unsupervised feature selection techniques. We illustrate as an example that when reducing the gene sets collections to a limited amount of pathways using unsupervised approaches, we can observe both increase or decrease a different number of significant pathways found on the selected association traits; on the other hand, whether the number of significant gene sets found is increasing or decreasing highly depends on the phenotypic trait and the gene sets collection used. 

\section{Discussion, limitations and future work}\label{sec:discussion}
Shapley values are often used to assign a fair value to players based on their contribution to the underlying game. However, they suffer from unawareness regarding redundancy among players which hinder their performance and can bias the resulting importance scores towards redundant players. In this work, we proposed a game-theoretical approach to incorporate redundancy-awareness into Shapley values to rank sets of genes in an unsupervised fashion. In particular, we proposed here different ways to punish the overlapping sets such that they are not progressively selected. 

%In our specific case, the players are sets of elements with potentially different dimensions and potential overlap. We showed that Shapley values alone are not sufficient for the ranking of players in this context as they do not consider the redundancy -- specifically the \emph{overlap} -- within the sets. Inspired by the microarray games presented by Moretti et al.~\cite{moretti_class_2007}, we illustrated our combination of Shapley values with redundancy-awareness in the context of gene sets and pathways. The notion of redundancy is highly dependent on the game-defining value function and on the players. In the context of gene pathways, players are sets of elements (pathways) and redundancy among two sets can be easily measured as the ratio among the size of the intersection and the one of their union. From here we derived the idea of using the Jaccard score to evaluate redundancy among sets and used it as punishment criteria in the rankings we proposed. 

We studied four different punishment methods and we were able to show that the orderings obtained are 
\begin{itemize}
    \item \emph{not favouring larger players} -- applying the redundancy-aware punishments avoids a positive correlation between the ranking and the size of gene sets
    \item \emph{redundancy free} -- the combination of Shapley values with the redundancy reduction criteria show high effectiveness in maintaining the importance of sets given by Shapley values while reducing the redundancy among players
    \item \emph{achieving high coverage} -- the obtained rankings still lead to high coverage of the genes. We showed that a positive correlation with the size of sets is not the unique solution to achieve high coverage of the genes, i.e., the original Shapley values ranking is not performing much better than the orderings which rank first small sets keeping low redundancy rates.
\end{itemize}

\subsection{Comparison among rankings}
In light of the results in Table~\ref{fig:jaccard_rate} and Table~\ref{fig:cumulative}, we conclude that AO leads to the worst ranking in covering the genes while the two re-scaled orderings together with SV are the best performing ones regarding coverage. When it comes to reducing the redundancy among pathways, all punished orderings can achieve this goal outperforming the SV ranking. Lastly, when comparing the different methods with respect to the correlation with the size of pathways, we see that only AOR and POR do not lead to any specific correlation. Hence, we conclude that POR is the best ordering one could choose when the aim was to optimize the ranking both for redundancy elimination and coverage of the genes without incurring specific correlations with the gene sets' sizes. 

\begin{table}[!t]
	\small
    \centering
    \newcolumntype{C}{>{\centering\arraybackslash}X}
    \begin{tabularx}{\textwidth}{@{}p{0.8cm}CCCCCC@{}}
    \toprule
    & \textbf{correlation with pathways' sizes} & \textbf{redundancy} & \textbf{coverage}  \\
    \midrule
        \textbf{SV}     & positive correlation              & reference level                   & reference level                   \\
        \textbf{PO}     &   negative correlation            & \textcolor{blue}{much less}     & \textcolor{blue}{same}    \\
        \textbf{POR}    & \textcolor{blue}{no correlation}  & \textcolor{blue}{less}          & less                         \\
        \textbf{AO}     &  negative correlation             & \textcolor{blue}{much less}     & \textcolor{blue}{same}  \\
        \textbf{AOR}    & \textcolor{blue}{no correlation}  & \textcolor{blue}{less}          & less                           \\
    
    \bottomrule
    \end{tabularx}
    \vspace*{1mm}
    \caption{\label{tab:summary}Comparison between the original SV and the newly proposed rankings.}
    \vspace{-2mm}
\end{table}
\begin{table}[!t]
	\small
    \centering
    \newcolumntype{C}{>{\centering\arraybackslash}X}
    \newcolumntype{R}{>1.5{\centering\arraybackslash}X}
    \begin{tabularx}{\textwidth}{@{}p{0.8cm}CCCCCCCC@{}}
    \toprule
    && {\textbf{SV}} & {\textbf{PO}} & {\textbf{POR}} & {\textbf{AO}} & {\textbf{AOR}}\\
    \midrule
    \textbf{KEGG} && 0.49 & -0.21 & 0.15 & 0.23 & 0.19 \\
    \textbf{CGN} && 0.43 & -0.51 & 0.019 & -0.702 & -0.041 \\
    \textbf{CM} && 0.759 & -0.531 & 0.019 & -0.702 & -0.041 \\
    \mbox{\textbf{TFT LEGACY} }&& 0.679 & -0.460 & 0.354 & -0.828 & 0.458 \\
    
    \bottomrule
    \end{tabularx}
    \caption{\label{fig:original_vs_bigsets} Kendall's $\tau$ coefficients measuring the correlation among the position in the ranking and the size of the gene sets.}
    \vspace{-2mm}
\end{table}
\subsection{Limitations and future work} We were able to show that using our tool as a pre-processing step for gene sets collections we get similar numbers of significant pathways when checking for association with phenotypic traits (although relying on much fewer pathways). We evaluated the significance of the top-ranked pathways with respect to phenotypic traits yielding very similar results to the unfiltered gene sets collections. However, our rankings based on Shapley values are unsupervised, they include only the structure of the gene sets collections and pathways while discarding the association among pathways and phenotypic traits; hence, we observe in some cases fewer significant pathways then considering the whole gene sets collections. 

%We strongly believe that our work can open to new research directions both in the applied as well i   n theoretical fields.

Our work could also lead to new perspectives for future research: One possible extension is the addition of a supervised punishment that considers also the relevance of each single pathway to a specific phenotypic trait. This could potentially lead to a higher number of significant pathways when using the Shapley values' based rankings to reduce the dimension of the gene sets collections. 

Furthermore, the proposed ranking methods  can be applied in any kind of family sets or where the relationship can be expressed as a binary matrix $B$; therefore, the possible applications of the proposed techniques are broad and various. From the theoretical point of view, a lot of research work can be done in understanding and creating a comprehensive theory on the redundancy among players and how to deal with the redundancy unawareness of Shapley values.

% _____________________________________________________________________________ 

\section{Conclusions}\label{sec:conclusions}
%\textbf{version 1:} We proposed a game theoretical interpretation of the problem where players are potentially overlapping sets. We developed four different methods to rank those sets, each of them satisfying different properties and achieving different goals. The main motivation for our work is the necessity and applicability of these rankings in applied research. In this work, we applied the selection of non-redundant and highly covering sets in family of sets in a case study on gene sets and pathways analysis. In this context, we were able to show very promising results on various metrics. However, one can assume that the range of potential applications is much broader. We strongly believe that our proposed methods can open new research paths both in the applied as well as in theoretical fields. 

We proposed the first redundancy-aware Shapley values to rank sets in families of sets. The four different presented rankings aim to satisfy various properties when selecting sets based on them. Motivated by the numerous applications of unsupervised feature selection, the proposed importance scores consider the distribution of elements within the family of sets and their overlap. In the presented application of the new methods to pathways in gene sets analysis, they showed favourable performances with respect to various metrics. However, one can assume that the range of potential applications is much broader. We strongly believe that our proposed methods can open new research paths both in the applied as well as in theoretical fields.

\bibliography{22geneset_biblio} 

\begin{thebibliography}{10}

\bibitem{agresti_introduction_2018}
{\sc A.~Agresti}, {\em An {Introduction} to {Categorical} {Data} {Analysis}},
  Wiley {Series} in {Probability} and {Statistics}, Wiley, 2018.

\bibitem{belinky_pathcards_2015}
{\sc F.~Belinky, N.~Nativ, G.~Stelzer, S.~Zimmerman, T.~Iny~Stein, M.~Safran,
  and D.~Lancet}, {\em {PathCards}: multi-source consolidation of human
  biological pathways}, Database, 2015 (2015).

\bibitem{benjamini_controlling_1995}
{\sc Y.~Benjamini and Y.~Hochberg}, {\em Controlling the {False} {Discovery}
  {Rate}: {A} {Practical} and {Powerful} {Approach} to {Multiple} {Testing}},
  Journal of the Royal Statistical Society. Series B (Methodological), 57
  (1995).

\bibitem{benjamini_control_2001}
{\sc Y.~Benjamini and D.~Yekutieli}, {\em The {Control} of the {False}
  {Discovery} {Rate} in {Multiple} {Testing} under {Dependency}}, The Annals of
  Statistics, 29 (2001).

\bibitem{berk_valid_2013}
{\sc R.~Berk, L.~Brown, A.~Buja, K.~Zhang, and L.~Zhao}, {\em Valid
  post-selection inference}, The Annals of Statistics, 41 (2013).

\bibitem{castro_polynomial_2009}
{\sc J.~Castro, D.~Gómez, and J.~Tejada}, {\em Polynomial calculation of the
  {Shapley} value based on sampling}, Computers \& Operations Research, 36
  (2009).

\bibitem{cohen_feature_2007}
{\sc S.~Cohen, G.~Dror, and E.~Ruppin}, {\em Feature selection via coalitional
  game theory}, Neural Computation, 19 (2007).

\bibitem{doderer_pathway_2012}
{\sc M.~S. Doderer, Z.~Anguiano, U.~Suresh, R.~Dashnamoorthy, A.~J. Bishop, and
  Y.~Chen}, {\em Pathway {Distiller} - multisource biological pathway
  consolidation}, BMC Genomics, 13 (2012).

\bibitem{dudoit_multiple_2008}
{\sc S.~Dudoit and M.~Laan}, {\em Multiple {Testing} {Procedures} {With}
  {Applications} to {Genomics}}, 2008.

\bibitem{fisher_logic_1935}
{\sc R.~A. Fisher}, {\em The {Logic} of {Inductive} {Inference}}, Journal of
  the Royal Statistical Society, 98 (1935).

\bibitem{frost_unsupervised_2016}
{\sc H.~R. Frost and C.~I. Amos}, {\em Unsupervised gene set testing based on
  random matrix theory}, BMC Bioinformatics, 17 (2016).

\bibitem{hochberg_sharper_1988}
{\sc Y.~Hochberg}, {\em A {Sharper} {Bonferroni} {Procedure} for {Multiple}
  {Tests} of {Significance}}, Biometrika, 75 (1988).

\bibitem{holm_simple_1979}
{\sc S.~Holm}, {\em A {Simple} {Sequentially} {Rejective} {Multiple} {Test}
  {Procedure}}, Scandinavian Journal of Statistics, 6 (1979).

\bibitem{jaccard_etude_1901}
{\sc P.~Jaccard}, {\em Etude de la distribution florale dans une portion des
  {Alpes} et du {Jura}}, Bulletin de la Societe Vaudoise des Sciences
  Naturelles, 37 (1901).

\bibitem{liberzon_molecular_2015}
{\sc A.~Liberzon, C.~Birger, H.~Thorvaldsdóttir, M.~Ghandi, J.~P. Mesirov, and
  P.~Tamayo}, {\em The {Molecular} {Signatures} {Database} ({MSigDB}) hallmark
  gene set collection}, Cell Systems, 1 (2015).

\bibitem{lucchetti_shapley_2010}
{\sc R.~Lucchetti, S.~Moretti, F.~Patrone, and P.~Radrizzani}, {\em The
  {Shapley} and {Banzhaf} values in microarray games}, Computers \& Operations
  Research, 37 (2010).

\bibitem{lundberg_unified_2017}
{\sc S.~Lundberg and S.-I. Lee}, {\em A {Unified} {Approach} to {Interpreting}
  {Model} {Predictions}}, arXiv:1705.07874 [cs, stat],  (2017).

\bibitem{mathur_gene_2018}
{\sc R.~Mathur, D.~Rotroff, J.~Ma, A.~Shojaie, and A.~Motsinger-Reif}, {\em
  Gene set analysis methods: a systematic comparison}, BioData Mining, 11
  (2018).

\bibitem{moretti_class_2007}
{\sc S.~Moretti, F.~Patrone, and S.~Bonassi}, {\em The class of microarray
  games and the relevance index for genes}, TOP, 15 (2007).

\bibitem{moretti_combining_2008}
{\sc S.~Moretti, D.~van Leeuwen, H.~Gmuender, S.~Bonassi, J.~van Delft,
  J.~Kleinjans, F.~Patrone, and D.~F. Merlo}, {\em Combining {Shapley} value
  and statistics to the analysis of gene expression data in children exposed to
  air pollution}, BMC Bioinformatics, 9 (2008).

\bibitem{nakagawa_farewell_2004}
{\sc S.~Nakagawa}, {\em A farewell to {Bonferroni}: {The} problems of low
  statistical power and publication bias}, Behavioral Ecology, 15 (2004).

\bibitem{noble_how_2009}
{\sc W.~S. Noble}, {\em How does multiple testing correction work?}, Nature
  Biotechnology, 27 (2009).

\bibitem{pfannschmidt_evaluating_2016}
{\sc K.~Pfannschmidt, E.~Hüllermeier, S.~Held, and R.~Neiger}, {\em Evaluating
  {Tests} in {Medical} {Diagnosis}: {Combining} {Machine} {Learning} with
  {Game}-{Theoretical} {Concepts}}, Information Processing and Management of
  Uncertainty in Knowledge-Based Systems, 610 (2016).

\bibitem{rozemberczki2022shapley}
{\sc B.~Rozemberczki, L.~Watson, P.~Bayer, H.-T. Yang, O.~Kiss, S.~Nilsson, and
  R.~Sarkar}, {\em The shapley value in machine learning}, 2022.

\bibitem{ref:shapley}
{\sc L.~S. Shapley}, {\em A value for n-person games}, Contributions to the
  Theory of Games,  (1953).

\bibitem{stoney_using_2018}
{\sc R.~A. Stoney, J.-M. Schwartz, D.~L. Robertson, and G.~Nenadic}, {\em Using
  set theory to reduce redundancy in pathway sets}, BMC Bioinformatics, 19
  (2018).

\bibitem{subramanian_gene_2005}
{\sc A.~Subramanian, P.~Tamayo, V.~Mootha, S.~Mukherjee, B.~Ebert, M.~Gillette,
  A.~Paulovich, S.~Pomeroy, T.~Golub, E.~Lander, and J.~Mesirov}, {\em Gene set
  enrichment analysis: {A} knowledge-based approach for interpreting
  genome-wide expression profiles}, Proceedings of the National Academy of
  Sciences of the United States of America,  (2005).

\bibitem{sun_game_2020}
{\sc M.~Sun, S.~Moretti, K.~Paskov, N.~Stockham, M.~Varma, B.~Chrisman,
  P.~Washington, J.-Y. Jung, and D.~Wall}, {\em Game theoretic centrality: a
  novel approach to prioritize disease candidate genes by combining biological
  networks with the {Shapley} value}, BMC Bioinformatics, 21 (2020).

\bibitem{van_campen_new_2017}
{\sc T.~van Campen, H.~Hamers, B.~Husslage, and R.~Lindelauf}, {\em A new
  approximation method for the {Shapley} value applied to the {WTC} 9/11
  terrorist attack}, Social Network Analysis and Mining, 8 (2017).

\bibitem{van_iersel_presenting_2008}
{\sc M.~P. van Iersel, T.~Kelder, A.~R. Pico, K.~Hanspers, S.~Coort, B.~R.
  Conklin, and C.~Evelo}, {\em Presenting and exploring biological pathways
  with {PathVisio}}, BMC Bioinformatics, 9 (2008).

\end{thebibliography}
\bibliographystyle{siam} 

\newpage
\section*{Appendix}
\subsection*{Sum-Of-Unanimity Games (SOUG)}\label{sec:SOUG_games} 
Consider a set of players $\mathcal{N}$ and a coalition $\mathcal{T}\subseteq \mathcal{N}$; the associated \emph{unanimity game} $u_{\mathcal{T}}$ is defined by
\begin{equation*}
    u_{\mathcal{S}}(\mathcal{T}) = 
    \begin{cases}
      1 \qquad \text{if } {\mathcal{S}} \subseteq {\mathcal{T}}\\
      0 \qquad \text{otherwise}
    \end{cases} \text{for any } {\mathcal{T}}\subseteq \mathcal{N}
\end{equation*}
It can be proved that given any cooperative game $(\mathcal{N},v)$, the value function $v$ can be written as the linear combination of Unanimity Games in a unique way, i.e., 
\begin{equation*}
    v(\cdot) = \sum_{{\mathcal{T}}\in \mathcal{P}(\mathcal{N})} \lambda_{\mathcal{S}}(v) u_{\mathcal{S}}(\cdot),
\end{equation*}
where $\lambda_{\mathcal{S}}(v)\in \mathbbm{R}$ are called \emph{unanimity coefficients} and are determined by the formula
\begin{equation*}
    \lambda_{\mathcal{S}}(v) = \sum_{{\mathcal{T}}\in \mathcal{P}(\mathcal{N})} (-1)^{s-t}v({\mathcal{T}})
\end{equation*}
As we see, the computation of $\lambda_{\mathcal{S}}(v)$, as well as the one of $\phi_i(v)$ becomes intractable if $N$ increases.

The SOUG allow for polynomial time computation of Shapley values. In particular, the computation in terms of the unaminity coefficients $\lambda_{\mathcal{S}}(v)$ is reduced to 
\begin{equation*}
    \phi_i(v)= \sum_{\mathcal{T}\subseteq\mathcal{N}\setminus \{i\}}\frac{\lambda_{\mathcal{S}}(v)}{|S|}
\end{equation*} for each player $i$ in $\mathcal{N}$. 

It can be proven that any cooperative game $(\mathcal{N},v)$ has a unique formulation as a sum of unanimity games. However, finding the equivalent SOUG of a game $(\mathcal{N}, v)$ is computationally equivalently hard as computing the Shapley values. 

Using SOUG brings the essential advantage of polynomial run-time when dealing with big families of sets, e.g., gene sets and pathways. 

\subsection*{Glove game}\label{sec:glovegame}
A classical example of a cooperative game is the so-called \emph{glove game}. Consider the set of players $\{A,B,C\}$; $A$ and $B$ are right-hand gloves while $C$ is a left-hand glove. A coalition, i.e., a subset of $\{A,B,C\}$, has value $1$ if it contains a pair of gloves (left + right) and has value $0$ if it does not. A person is wearing one pair of gloves at a time, therefore adding more gloves to a coalition already containing a pair of gloves is useless; we represent this mathematically - any coalition containing a pair does not increase its worth when including more gloves. The \emph{grand coalition} $\{A,B,C\}$ contains one pair of gloves, i.e., the pair $\{A,C\}$ or the pair $\{A,B\}$, therefore it has value equals to $1$. Note that the value function assigns $1$ to the grand coalition and $0$ to the empty set. After computing the Shapley values, we find $\phi(A) = \phi(B) = \frac{1}{6}$ and $\phi(C) = \frac{1}{3}$. Both players $A$ and $B$ get the same Shapley values as they are essentially indistinguishable. Shapley values scores do not detect the existing \emph{redundancy} among $A$ and $B$. After including one element among $A$ and $B$, including the other does not yield any advantages. We refer to this similarity among players as \emph{redundancy} and we say that the Shapley values are unaware of redundancy among players.

\subsection*{Example of computation of Shapley values}

We give here an example on how to compute the Shapley values in a toy example. Consider a family of sets 

\begin{equation*}
\mathcal{F} = \{P_1 = \{a,b,c,e\}, P_2= \{a,d\}, P_3= \{c,d,e\}\}.
\end{equation*}

The union of the sets in $\mathcal{F}$ is $G= \{a,b,c,d,e\}$. We construct the binary matrix $B$, i.e.,
\begin{equation*}
    B= 
    \left(
    \begin{array}{ccccc}
         1 & 1 & 1 & 0 & 1\\
         1 & 0 & 0 & 1 & 0\\
         0 & 0 & 1 & 1 & 1
    \end{array}
    \right)
\end{equation*}
where each column represents an elements of $G$, respectively $a,b,c,d,e$ and the zeroes and ones components of $B$ represent the binary relationship of being included in each set of $\mathcal{F}$. We then get the dictionary $\mathcal{A}$ as previously described. From the first column of $B$ follows that $a$ belongs to $P_1$ and $P_2$ but not to $P_3$ thus the set $\{P_1,P_2\}$ is the first set in $\mathcal{A}$. After applying the same procedure on each column of $B$, we get 
\begin{equation*}
    \mathcal{A} = \{\{P_1, P_2\}, \{P_1\}, \{P_1, P_3\}, \{P_2, P_3\}, \{P_1, P_3\}\}.
\end{equation*}
Now, we calculate the Shapley values as in Equation~\eqref{eq:shapley_definition}, leading to
\begin{align*}
    &\phi(P_1) = \frac{1}{5}\cdot\left(1\cdot\frac{1}{2}+1\cdot1+1\cdot\frac{1}{2}+0\cdot\frac{1}{2}+1\cdot\frac{1}{2}\right) =\frac{1}{2}\\
    &\phi(P_2) =\frac{1}{5}\cdot\left(1\cdot\frac{1}{2}+0\cdot 1+0\cdot\frac{1}{2}+1\cdot\frac{1}{2}+0\cdot\frac{1}{2}\right) = \frac{1}{5}\\
    &\phi(P_3) = \frac{1}{5}\cdot\left(0\cdot\frac{1}{2}+0\cdot1+1\cdot\frac{1}{2}+1\cdot\frac{1}{2}+1\cdot\frac{1}{2}\right) = \frac{3}{10}
\end{align*}
As expected, we observe that the sum of Shapley values equals $1$. Moreover, we can notice that there is a bias towards sets that have higher dimensions, i.e., the ordering $P_1$, $P_3$ and $P_2$ reflects also the ordering of the sets with respect to their sizes. This example illustrates that Shapley values do not aim for low intersection among ranked sets or high coverage.\par
The sets in $\mathcal{F}$ are ordered according to Shapley values as $P_1, P_3$ and $P_2$. However $P_1$ and $P_3$ share two elements while $P_1$ and $P_2$ share only one. If we want to select two out of these three subsets in order to maximize the coverage of $G$ while keeping low the overlapping among the sets, we would better select $P_1$ and $P_2$ instead of $P_1$ and $P_3$. 

\subsection*{Different rankings}

We introduced in the paper four different orderings which depend on different punishment criteria. The difference among the punishments criteria is reflected in the coverage and redundancy reduction performances of the various orderings. We give here detailed information on the algorithm for sake of completeness and for reproducibility purposes. \par 
The Shapley values need to be re-computed after each iteration as, after the selection of some pathways, they do not sum anymore to $1$.
\begin{itemize}
    \item \textbf{Punished Ordering (PO)} --  Moreover, after selecting one pathway, the order of importance can be potentially changed. 
    \item \textbf{Punished Ordering Re-scaled (POR)} -- tThe algorithm shows strong similarities with PO. The difference among the two rankings is in the punishment: the punishment implemented in PO is re-scaled in POR. The complexity of the algorithm proposed does not change.
    
    This avoids that, after the first $n$ steps, the highest importance scores assume negative real numbers as the punishments can assume values higher than $1$. 
    The importance scores are defined as follows:
    \begin{equation*}
        \begin{array}{lcl}
            S_1(P) &= &\phi_1(P)\\            
            R_n &= &\frac{\max_P \phi_n(P)}{\max_P \sum_{i=1}^n j(\arg\max_{\bar{P}} S_i(\bar{P}), P)}\\
            S_{n+1}(P)&=& \phi_{n+1}(P) - \sum_{i=1}^n j(\arg\max_{\bar{P}} S_i(\bar{P}), P)\cdot R_n, \qquad \text{if } n\geq 0.
        \end{array}
    \end{equation*}
    \item \textbf{Artificial Ordering (AO)} -- Still based on the original Shapley value ordering. After ranking the first pathway, it computes an \emph{artificial pathway} which includes all genes which are including at least in one of the pathways previously selected. 
    
    In order to rank the $(n+1)$-th set, PO or POR punish multiple times the overlap among sets that share some element $g$. The introduction of the artificial gene set $AP_n$ avoids these multiple punishments.
    \begin{equation*}
        \begin{array}{lcl}
            S_1(P) &= &\phi_1(P)\\
            S_{n+1}(P)&=& \phi_{n+1}(P) - j(AP_n, P), \qquad \text{if } n\geq 0.
        \end{array}
    \end{equation*}
    \item\textbf{Artificial Ordering Re-scaled (AOR)} -- The difference with the AO ordering is the re-scaling of the punishment terms. The complexity of the algorithm proposed does not change.
    \begin{equation*}
        \begin{array}{lcl}
            S_1(P) &= &\phi_1(P)\\  
            R_n &= &\frac{\max_P \phi_n(P)}{\max_P j(AP_n, P)}\\
            S_{n+1}(P)&=& \phi_{n+1}(P) - j(AP_n, P_j)\cdot R_n, \qquad \text{if } n\geq 0.
        \end{array}
    \end{equation*}

\end{itemize}

\subsection*{Data}\label{sec:data}
We apply our methods on gene sets, i.e., lists of genes based on a potential common biological functionality. Typically, the gene sets are grouped in gene sets collections according to some a-priori biological knowledge, e.g., involvement in the same biological processes, presence of common biochemical mechanisms or also shared associations with a phenotype~\cite{liberzon_molecular_2015}. We selected four gene sets collections, i.e., KEGG, CGN, CM and TFT LEGACY, from the GSEA library\footnote{http://www.gsea-msigdb.org/gsea/msigdb/ns.jsp\#C5} and some clinical traits\footnote{http://twas-hub.org/traits/} in order to assess whether associations exist between gene sets and traits. 
Each selected gene set collection contains a number of pathways ranging from $186$ to $610$; the association traits considered contain at least $500$ relevant genes.

\subsection*{Availability of data and materials}
The datasets supporting the conclusions of this article are available at the \href{http://www.gsea-msigdb.org/gsea/msigdb/collections.jsp}{GSEA repository}. The implementation code is available on \href{https://github.com/chiarabales/geneset_SV}{Github}.

\end{document}